\documentclass[aps,prl,showpacs,twocolumn]{revtex4}
\usepackage{graphicx}
\input{epsf}

\begin{document}
\title {Quantum cavity modes in spatially extended
Josephson systems}
\author{M. V. Fistul$^{1}$ and A. V. Ustinov$^{2}$}

\affiliation {$^{1}$ Theoretische Physik III, Ruhr-Universit\"at
Bochum, D-44801 Bochum, Germany \\
$^2$Physikalisches Institut III, Universit\"at
Erlangen-N\"urnberg, D-91058, Erlangen, Germany }

\date{\today}
\begin{abstract}
We report a theoretical study of the macroscopic quantum dynamics
in spatially extended Josephson systems. We focus on a Josephson
tunnel junction of finite length placed in an externally applied
magnetic field. In such a system, electromagnetic waves in the
junction are excited in the form of cavity modes manifested by
Fiske resonances, which are easily observed experimentally. We
show that in the quantum regime various characteristics of the
junction as its critical current $I_c$, width of the critical
current distribution $\sigma$, escape rate $\Gamma$ from the
superconducting state to a resistive one, and the time-dependent
probability $P(t)$ of the escape are influenced by the number of
photons excited in the junction cavity. Therefore, these
characteristics can be used as a tool to measure the quantum
states of photons in the junction, e.g. quantum fluctuations,
coherent and squeezed states, entangled Fock states, etc.
\end{abstract}


\maketitle

Great interest is currently attracted to experimental and
theoretical studies of macroscopic quantum phenomena in diverse
Josephson systems \cite{Tinkham,Clarke1,Martinis-Lukens-et-al}.
Most of such systems contain just one or few {\it lumped}
Josephson junctions. At low temperatures, quantum-mechanical
effects such as macroscopic quantum tunneling, energy level
quantization, and coherent oscillations of the Josephson phase
have been observed \cite{Clarke1,Martinis-Lukens-et-al}. The
important method allowing to study the macroscopic energy levels
in Josephson coupled systems is the microwave spectroscopy
\cite{Clarke1,Martinis-Lukens-et-al}. As the frequency $\omega$ of
an externally applied microwave radiation matches the energy level
separation, one can observe resonant absorption and Rabi
oscillations due to the population of excited levels.

An interaction between Josephson systems and \emph{quantized
electromagnetic fields} opens new frontiers in research. Similar
type of interaction arises between an atom and electromagnetic
cavity modes in quantum electrodynamics (QED). Recently, such
QED-inspired experiments have been performed with superconducting
charge qubits \cite{Wallraff-Nature-2004} and flux qubits
\cite{Chiorescu-Nature-2004,Johansson-PRL-2006} coupled to on-chip
resonators. In these experiments, the quantum degree of freedom of
a lumped Josephson system couples to the quantized electromagnetic
field of a superconducting cavity located on the same chip, as
illustrated in Fig.~1(a). The excitation of \emph{cavity modes}
CMs in an external cavity leads to an appearance of an oscillating
current flowing through the Josephson circuit. Such an oscillating
current excites transitions between macroscopic energy levels of
the Josephson phase. Quantum regime of a weak resonant interaction
between the Josephson phase $\varphi$ and CMs is described by a
bilinear Hamiltonian $\hat H~\propto~\varphi (\hat a +\hat a^+)$,
where $\hat a^+ (\hat a^) $ are the operators of creation
(annihilation) of a particular CM. This Hamiltonian corresponds to
the famous Jaynes-Cummings model \cite{Loudon,BBB} and therefore,
a quantum regime of a lumped Josephson circuit incorporated in an
external transmission line can be mapped to a problem of a single
atom weakly interacting with CMs. In this regime, many fascinated
phenomena as mixture of different Rabi frequencies, creation of
entangled states of CMs, a single atom maser behavior, can be
observed \cite{Girvin}. The interaction via CMs can be used to
couple superconducting qubits \cite{Makhlin,Zag,FistUstInt}.

As we turn to {\it spatially extended} Josephson systems, e.g.
long Josephson junctions, Josephson junction parallel arrays and
ladders, the spatially-dependent Josephson phase $\varphi(x)$ can
also display macroscopic quantum effects as tunneling and energy
level quantization. For a Josephson vortex trapped in a long
Josephson junction, these phenomena have been studied
theoretically \cite{KIM} and observed in experiments
\cite{Wallraff03}. An additional, interesting property of
spatially extended junctions is that they can support the
propagation of electromagnetic waves forming CMs \emph{inside} the
junction itself, see Fig.~1(b). When the Josephson system is
biased in the resistive state the resonant interaction of such CMs
with ac Josephson current leads to the classical resonances in the
current-voltage characteristics known as Fiske steps
\cite{Fiske,Cirillo-1998}. The quantum regime of such resonances
at finite voltages in systems containing few small Josephson
junctions has been theoretically considered in
Ref.~\cite{FistPSS}. The next step naturally appearing in a study
of macroscopic quantum phenomena in spatially extended Josephson
systems is the quantum regime of interaction between the Josephson
phase and \emph{intrinsic CMs} of the system.

\begin{figure}
\noindent \includegraphics[width=3.5in]{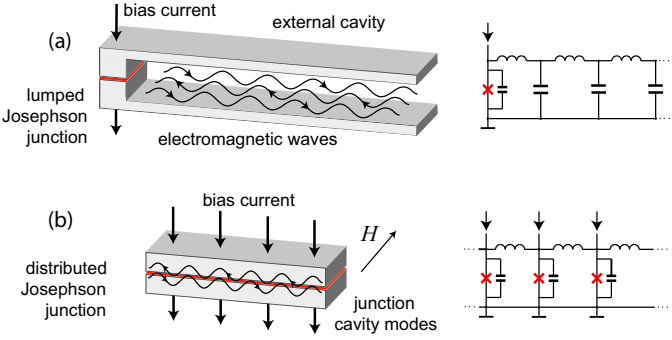} \caption{(a)
Schematic view of lumped Josephson junction embedded in an
external cavity and its equivalent scheme. (b) Distributed
Josephson junction subject to an external magnetic field with
excited internal (Fiske) CMs and its equivalent scheme.}
\label{sketch}
\end{figure}

In a current-biased Josephson junction, the spatially-averaged
Josephson phase difference $\varphi$ oscillates at Josephson
plasma frequency
$\omega_p=\omega_{p0}\left[1-(I/I_{c0})^2\right]^{1/4}$ depending
on the dc bias $I$. Here, $\omega_{p0}$ and $I_{c0}$ are the
plasma frequency at zero bias and the nominal value of the
critical current of the junction, respectively. If the Josephson
plasma frequency $\omega_p$ matches the frequency of CMs, the
interaction between the Josephson phase and CMs becomes resonant.

In this paper, we consider the alternative, non-resonant
interaction between the Josephson phase $\varphi$ and intrinsic
CMs of the junction. Such an interaction can be realized in long
Josephson junctions or their discrete versions -- parallel
Josephson arrays, often also called as Josephson transmission
lines. We show that the switching from the superconducting state
to a resistive one, i.e. the Josephson phase escape can be
facilitated or suppressed by excitation of cavity modes.

Moreover, we argue that measurements of the switching current
$I_c$, the width of its distribution $\sigma$, and the
time-dependent probability $P(t)$ of the escape allow to obtain
information on the {\it quantum state} of intrinsic CMs in the
junction.

In order to quantitatively analyze the macroscopic quantum
phenomena appearing due to the interaction of plasma oscillations
with CMs, we consider a Josephson junction of a finite length $L$.
The Josephson junction is characterized by the time and coordinate
dependent Josephson phase $\varphi(x,t)$. In the presence of an
externally applied magnetic field the Josephson phase can be
written as
\begin{equation} \label{JP-MagField}
\varphi(x,t)~=~\frac{2\pi \Phi x}{\Phi_0 L}
+\tilde{\varphi}(x,t)~~,
\end{equation}
where $\Phi$ is the external magnetic flux, and $\Phi_0=hc/2e$ is
the flux quantum \cite{Comment-JL}. The system is described by
Hamiltonian:
$$
H=E_{J0}\int_0^L \Biggl [
\frac{1}{2\omega_{p0}^2}\tilde{\varphi}_t^2
+\frac{\lambda_J^2}{2}\tilde{\varphi}_x^2+U(x,\tilde{\varphi})
\Biggr ]\frac{dx}{L}~,
$$
\begin{equation} \label{Hamilt}
U(x,\tilde{\varphi})~=~E_{J0} \Biggl  [1-\cos\left(\frac{2\pi \Phi
x}{\Phi_0 L} +\tilde{\varphi}\right) -i\tilde{\varphi} \Biggr ] ~,
\end{equation}
where $\lambda_J$ and $E_{J0}$ are the Josephson penetration
length and the Josephson energy, accordingly. The normalized dc
bias $i=I/I_{c0}$ can be changed to tune the effective potential
$U(x,\tilde{\varphi})$ of the junction.

Next, we represent the time-dependent Josephson phase by a sum of
electromagnetic CMs and a "center mass" Josephson phase $\chi(t)$,
i.e.
\begin{equation} \label{Sum:Josphase}
\tilde{\varphi}(x,t)~=~\chi(t)+\sum_{n} Q_n(t) \cos(k_n x),
\end{equation}
where the wave numbers of cavity modes $k_n=\pi n/L$, with
$n=1,2...$. Substituting expression (\ref{Sum:Josphase}) in Eq.
(\ref{Hamilt}) and assuming that the amplitudes of cavity modes
$Q_n(t)$ are small (so that we can neglect anharmonic interaction
between CMs \cite{comment}), we obtain the Hamiltonian in the
following form:
$$
H=\frac{E_{J0}}{2\omega_{p0}^2} \Biggl [
\dot{\chi}^2+\frac{1}{2}\sum_{n} \dot{Q}_n^2
+\frac{\omega_n^2}{2}Q_n^2 \Biggr]+ U(\chi,Q_n),
$$
\begin{equation} \label{Hamiltonian:q}
U(\chi,Q_n)~=~E_{J0} \sum_{n,m} a_{nm} Q_n Q_m-E_J(H)\cos \chi
-i\chi~,
\end{equation}
where $\omega_n=\omega_{p0}\lambda_J k_n$. The magnetic field
dependent coupling energy is $E_J(H)=E_{J0}\frac{I_c(H)}{I_{c0}}$,
where $I_c(H)$ is the magnetic field suppressed nominal value of
the critical current, i.e.
$$
I_c(H)=I_{c0}\left|\frac{\sin(\frac{\pi \Phi}{ \Phi_0}) } {
\frac{\Phi}{\Phi_0} } \right|~~.
$$
The $\chi$-dependent coefficients
$$
a_{nm}~=~\frac{1}{2}\int \frac{dx}{L} \cos(k_n x) \cos(k_m x) \cos
\left(\frac{2\pi \Phi x}{\Phi_0 L} +\chi\right)
$$
determine the strength of interaction between the CMs and the
center mass Josephson phase.

As $E_J(H)\gg\hbar\omega_{p0}$, the switching from the
superconducting state to a resistive one occurs at the dc bias
close to its critical value, i.e. $\delta~=~[I_c(H)-I]/I_c(H)\ll~1
$, and the quantum-mechanical Hamiltonian (\ref{Hamiltonian:q})
can be written in the following form:
\begin{equation} \label{Hamilt:total}
 \hat H ~=~\hat H_0 +\sum_{n} \hat H_n +\hat H_{int}~~,
\end{equation}
where the first term
\begin{equation} \label{FirstHam}
 H_0~=~\frac{\omega_{p0}^2}{2E_{J0}}\hat P_{\chi}^2+E_J(H)\left[\delta
 \chi-\frac{\chi^3}{6}\right]
\end{equation}
describes the dynamics of homogeneous (plasma) oscillations of the
Josephson phase with momentum operator $\hat P_{\chi}$. The second
term is the Hamiltonian of noninteracting CMs
\begin{equation} \label{SecHam}
 H_n~=~\frac{\omega_{p0}^2 }{E_{J0}}\hat P_{Q_n}^2+
 E_{J0}\frac{\omega_n^2}{4\omega_{p0}^2}Q_n^2~,
 \end{equation}
where $\hat P_{Q}=-i\hbar \frac{\partial}{\partial Q}$ is the
operator of generalized momentum . Here, we have neglected a small
renormalization of CMs spectrum due to the presence of plasma
oscillations of the Josephson phase $\chi$.
The last term in (\ref{Hamilt:total}) describes an interaction
between the Josephson phase and CMs:
\begin{equation} \label{Hamilt:interaction}
\hat H_{int} ~=~-E_J\chi \sum_{n } a_n Q_n^2~,
\end{equation}
where $a_n=\frac{1}{2}\int \frac{dx}{L} \cos^2(k_n x) \cos
(\frac{2\pi \Phi x}{\Phi_0 L})$. In the absence of an external
magnetic field all coefficients $a_n=1/4$. However, in the
presence of a magnetic field characterizing by magnetic flux,
$\Phi \simeq m \Phi_0$, the coefficient $a_m~=-1/8$, and other
coefficients are small.

If the size of the Josephson system is not very large
($L\leq\lambda_J$), all frequencies $\omega_n\gg\omega_{p0}$ and,
therefore, the interaction between the CMs and the plasma
oscillations is non-resonant. Thus we can use a time-averaged
expression for the interaction Hamiltonian, $\hat H_{int} =E_J\chi
\sum_{n} a_n<Q_n^2>$, where $<...>$ is the time average. The mean
switching current $\bar I_c$ is written as
\begin{equation} \label{CriticalCurr}
\bar I_c=I_c(H)-I_{c0}\sum_{n}a_n<Q_n^2>~~.
\end{equation}
The excitation of CMs results in a nonzero value of $<Q_n^2>$,
which leads to a {\it suppression} of the average switching
current in the absence of magnetic field. However, when the
magnetic field is applied, the fluctuation-free critical current
$I_c(H)$ is strongly suppressed and excitation of CMs results in
an unusual \emph{increase} of the mean switching current ($a_n$
values become negative in this case).

Switching from the superconducting (zero-voltage) state to the
resistive state occurs at random values of the external current
$I$. This process is characterized by the probability $P(I)$. If
we neglect the CMs, the Josephson phase escape in the quantum
regime is determined by macroscopic tunneling, and the width
$\sigma$ of the distribution $P(I)$ strongly depends on magnetic
field:
\begin{equation} \label{widthDistr:noCMs}
\sigma~\simeq~\left(\frac{\hbar \omega_{p0}}{E_{J0}}\right)^{4/5}
I_{c0}^{2/5}I_c(H)^{3/5}~~.
\end{equation}
Quantum fluctuations induced by zero-point oscillations of CMs
lead to a {\it saturation} of $\sigma$ as a function of magnetic
field at the level
\begin{equation} \label{widthDistr:CMs}
\sigma_{CM}~\simeq~I_{c0}\frac{\hbar \omega_{p0} L}{E_{J0}
\lambda_J}~.
\end{equation}
The typical dependence of the width of the critical current
distribution on magnetic flux applied to the junction is shown by
solid line in Fig.~2.

\begin{figure}
\includegraphics[width=2.4in]{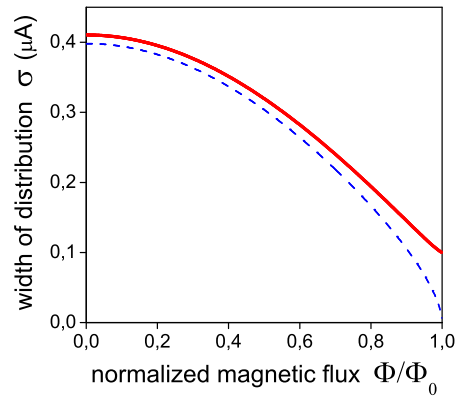}
\caption{The width of the switching current probability
distribution as a function of magnetic field with (solid line) and
without (dashed line) taking into account zero-point oscillations
of CMs. Here we have taken typical parameters for the junction as
$I_{c0}=100 \mu A$, and $\hbar \omega_{p0}/E_{J0}=10^{-3}$, the
junction size $L~= \lambda_J$.} \label{widthP}
\end{figure}

In the following, we present a quantitative analysis of the
influence of quantum-mechanical properties of CMs on the process
of escape from the superconducting state to the resistive state of
the junction. In the quantum regime the process of escape is
determined by the tunneling of phase $\chi$ and the switching rate
is given by \cite{Tinkham,Clarke1,Martinis-Lukens-et-al}
\begin{equation} \label{Gamma1}
 \Gamma (I,Q_n)\propto e^{ -\frac{48 (\sqrt{2}E_J(H)E_{J0})^{1/2}}
 {5\hbar \omega_{p0}}\left(\delta -\frac{I_{c0}}{I_c(H)}
 \sum_{n}a_n Q_n^2\right)^{5/4}}.
\end{equation}
In experiment, the measured characteristics is usually the
time-dependent probability $P(I,t)$ of finding the junction in the
zero-voltage state. Since, $Q_n$ are the random variables, this
probability can be written as
\begin{equation} \label{P1}
P(I,t) = \int \rho(Q_n) \exp\left[-t\,\Gamma(I,Q_n)\right]dQ_n ~.
\end{equation}
Here, $\rho(Q_n)$ is the quantum-mechanical probability
distribution of CMs in the junction.

If we neglect the excitation of CMs, $P_0(t) \simeq
\exp(-t/\tau_0)$ displays a standard exponential decay with time.
Here, $\tau_0=1/\Gamma(I,0)$ is the dc bias dependent lifetime of
the zero-voltage state, see Eq. (\ref{Gamma1}). This behavior is
shown in Fig. 3 by the dashed line. The deviation of $P(t)$ from
the exponential dependence allows to characterize the
quantum-mechanical properties of CMs. The situation becomes
especially interesting for {\it nonequilibrium and nonclassical
states} of the cavity modes. These are the zero-point
oscillations, the chaotic, coherent and squeezed states, and
various entangled states well known in quantum optics
\cite{Loudon,BBB}.

As a particular example, first, we consider zero-point
oscillations in CMs. In this case the $\rho(Q_n)$ is given as
\begin{equation} \label{Quantumnoise:denmatr}
 \rho_{\rm zp}(Q_n)=\prod_n \left(\frac{E_{J0}\omega_n}{2\pi \hbar´\omega_{p0}^2}\right)^{1/2}
 \exp \left(-\frac{E_{J0}}{2\omega_{p0}^2\hbar}
 \sum_{n}\omega_n
 Q_n^2 \right)~~.
\end{equation}
Substituting (\ref{Quantumnoise:denmatr}) in Eq. (\ref{P1}) and
calculating all integrals over $Q_n$, we obtain for
$\Phi\approx\Phi_0$
\begin{equation} \label{P:quantumnoise}
 P_{CM}^{\rm zp}(t)\simeq \left(\frac{\gamma_1 \tau_0}{t}\right)^{\gamma_1},
 ~~t \gg \gamma_1 \tau_0,
\end{equation}
where the magnetic field dependent parameter $\gamma_{1}=\frac{\pi
\lambda_J}{3 L
}\left(\frac{I_c(H)}{I_{c0}}\right)^{1/2}(2\delta)^{-1/4}$. For
small times $t<\gamma_1\tau_0$ the influence of the equilibrium
CMs quantum noise is small, and the dependence
$P(t)\simeq\exp(-t/\tau_0)$ restores. Notice here, that for small
external magnetic field and not very large number of excited CMs
the exponential dependence of $P(I,t)$ is always valid.

Various nonequilibrium quantum-mechanical states of CMs can be
induced by applying external microwaves. The interesting case is
the {\it coherent state} of a single CM excited at a frequency
$\omega=\omega_1$. In this case all values of $Q_n$ except for
$Q_1$ are small, and  $\rho(Q_1)$ is given by
\begin{equation} \label{Densmatr:coherentstate}
 \rho_{coh}(Q_1)\simeq\exp \left[-\frac{E_J\omega_1}{2 \omega_{p0}^2\hbar}(Q_1-\eta)^2
 \right]~~,
\end{equation}
where $\eta\propto\sqrt{W}$ is determined by the power $W$ of
external microwave radiation. For a Josephson junction subject to
an external magnetic field, we obtain that the excitation of the
coherent state of a single CM leads to the dependence $P(t)$ in
the form
\begin{equation} \label{P:coherentstate}
 P_{CM}^{\rm coh}(t)~\simeq~\exp
 \left\{-\gamma_{2}\ln^2 \frac{t}{\gamma_{2}\tau_{\rm coh}} \right
 \},~~t\gg\gamma_{2}\tau_{\rm coh}.
\end{equation}
The parameter $\gamma_{2}~\simeq~\frac{ \lambda_J}{L }\frac{\hbar
\omega_{p0}I_c(H)}{E_{J0} I_{c0}\eta}\delta^{-1/2}$ depends on
both magnetic field and the power of microwave radiation. Here
$\tau_{\rm coh}=1/\Gamma (I,Q_1=\eta)$, see Eq.~(\ref{Gamma1}).
Similarly to the case of a quantum noise in CMs, at short times
$t<\gamma_{2}\tau_{\rm coh}$ the $P(I,t)$ decays exponentially
with time, i.e. $P_0(t)\simeq\exp(-t/\tau_{\rm coh})$.

Another interesting quantum-mechanical state of CMs is the chaotic
state of a single CM induced by microwave radiation
\cite{Loudon,BBB}. In this case the power $W$ of external
microwave radiation determines only the mean photon number
$\bar{m}$, i.e. $\bar{m}~\propto~ W$. As the mean photon number
$\bar{m}$ is relatively large the probability distribution
$\rho(Q_n)$ takes the form
\begin{equation} \label{CSCM}
 \rho_{\rm ch}(Q_n)=\left(\frac{E_{J0}\omega_1}{2\pi \omega_{p0}^2\hbar}\bar{m} \right)^{1/2}
 \exp \left(-\frac{E_{J0}\omega_1}{2\hbar\omega_{p0}^2 \bar{m}}
Q_1^2 \right)~~
\end{equation}
Therefore, a strongly excited chaotic state of a single CM should
show the same time-dependent probability $P(t)$ as in the
zero-point oscillations case (\ref{P:quantumnoise}), for which
$\gamma_1$ is replaced by the microwave power dependent parameter
$\gamma_3=\gamma_1/\bar{m}$.

The probability $P(t)$ of finding the junction in the zero-voltage
state is shown in Fig.~3 for various states of CMs. We note that
more complicated states as squeezed states or Fock states (e.g.,
$N$ photons in one mode $n=1$) can be prepared by using pulsed
technique and intrinsic nonlinearity of cavity modes \cite{Yurke}.
The entanglement of Fock states will manifest itself by
oscillations of $P(t)$ dependence. For realistic values of
junction parameters we obtain $\gamma_1\simeq\gamma_2=7$, and the
deviations of $P(t)$ from the exponential decay should be
detectable experimentally.

\begin{figure}
\includegraphics[width=2.4in]{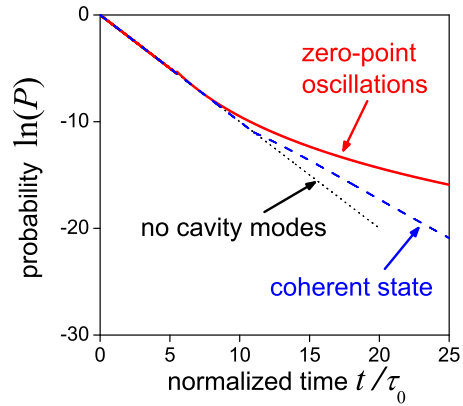}
\caption{Time-dependent probability $P(t)$ of finding the junction
in the zero-voltage state: without taking into account CMs (dotted
line) and with taking into account quantum fluctuations induced by
CMs (solid line) and coherent state of CMs (dashed line). The
parameters are chosen as $\delta=5~10^{-3}$, $I_c(H)/I_c(0)=0.2$,
$\hbar \omega_{p0}/E_{J0}=10^{-4}$, $\eta=1.5 ~10^{-4}$ and the
junction size $L~=0.27 \lambda_J$. For simplicity, we have taken
$\tau_0=\tau_{\rm coh}$.} \label{Pdependence}
\end{figure}

In conclusion, we have shown that the excitation of cavity modes
in distributed Josephson junction or parallel arrays of junctions
manifests itself by either enhancement or suppression of the
escape rate from the superconducting state, depending on applied
magnetic field. This effect is due to a renormalization of the
potential barrier for the escape which, in turn, depends on the
quantum state of the cavity mode. The important characteristics of
the cavity mode quantum electrodynamics, namely the probability
distribution of the CMs $\rho(Q)$ can be detected experimentally
by measuring the temporal decay of the switching probability
$P(t)$ given by Eqs.~(\ref{Gamma1}) and (\ref{P1}).

M. V. F. acknowledges the financial support by SFB 491.


\begin{references}
\bibitem{Comment-JL} Strictly speaking the Eq. (\ref{JP-MagField})
is valid in the limit of $L \ll \lambda_J$, where $\lambda_J$ is
the Josephson penetration length. However, our analysis is
qualitatively correct also for junctions of a size $L \simeq
\lambda_J$ and subjected to large magnetic field when the critical
current is strongly suppressed.
\bibitem{comment} In particular, we neglect here by an interaction between the
cavity modes with different wave numbers. Such an interaction can
be used in order to prepare various quantum states of Josephson
cavity modes \cite{Yurke}.
\end{references}
\end{document}